\documentclass[conference]{IEEEtran}
\usepackage{cite}
\usepackage{amsmath,amssymb,amsfonts}
\usepackage{graphicx}
\usepackage{diagbox}
\usepackage{textcomp}
\usepackage{xcolor}
\usepackage{url}
\def\BibTeX{{\rm B\kern-.05em{\sc i\kern-.025em b}\kern-.08em
    T\kern-.1667em\lower.7ex\hbox{E}\kern-.125emX}}
\begin{document}

\title{Continuous data acquisition for liquid argon time projection chamber neutrino detectors using FPGA-based real-time compression algorithms}

\author{\IEEEauthorblockN{J. I. Crespo-Anad\'on, for the MicroBooNE Collaboration}
\IEEEauthorblockA{
\textit{Columbia University}\\
New York, NY, 10027, USA\\
jcrespo@nevis.columbia.edu}
}

\maketitle

\begin{abstract}
Liquid argon time projection chambers (LArTPCs) have been proposed as neutrino detectors that combine both large sizes to maximize the number of neutrino interactions and detailed recording of the interaction. The readout of thousands of channels at MHz sampling rates produce images of the neutrino-nucleus interaction with millimeter-scale resolution, enabling the identification of the resulting particles and offering multiple handles to measure their energies at the expense of large data rates. Continuous acquisition of the LArTPC data is required to enable the use of high-latency triggers for capturing non-accelerator-beam events. We describe the case of the continuous readout of the MicroBooNE LArTPC, that grants the possibility of acquiring the neutrino burst from a supernova using the SNEWS (Supernova Early Warning System) alert as delayed trigger. The continuous data acquisition is accomplished by using real-time compression algorithms (zero suppression and Huffman compression) implemented in an FPGA in the readout electronics.
\end{abstract}


\section{Introduction}\label{Introduction}
Neutrinos are among the least understood elementary particles. The observation of neutrino oscillations, the periodic change of neutrino flavor as they propagate, reveals that neutrinos have non-zero mass, but their masses and ordering are unknown. In addition, it is not known if neutrinos violate the charge-parity (CP) symmetry, which would introduce fundamental differences between neutrino and antineutrino oscillations. Moreover, there are some experimental anomalies which can be understood as the existence of additional neutrinos beyond the 3 neutrinos of the Standard Model which require further investigation \cite{Aguilar-Arevalo:2018gpe}.

A controlled source of neutrinos is provided by particle accelerators. Protons are accelerated and made to collide against a fixed target, resulting in unstable charged mesons, 
producing neutrinos when they decay.
This source has the advantage of providing a neutrino beam with a known timing, which can be used to open the data-acquisition window in the detector.
There are other sources of neutrinos which do not feature a known timing, which require a different approach for acquiring them. Among them, the neutrino burst from a core-collapse supernova is especially interesting, as it carries information about neutrino physics and the explosion dynamics and ultimate fate of the star.

\section{The MicroBooNE detector}\label{MicroBooNE}
The MicroBooNE detector \cite{Acciarri:2016smi} is a liquid argon time projection chamber (LArTPC) located in the Booster Neutrino Beam at Fermilab, taking neutrino beam data since October 2015. It has an active mass of 85 t, with a maximum electron drift time of 2.3 ms. Three wire planes with a 3-mm wire pitch are used to detect the ionization electrons produced by interacting particles in the liquid argon: 2 induction planes with 2400 wires each at $\pm 60^\circ$ from the vertical, and a collection plane with 3256 vertical wires. In addition, 32 PMTs located behind the wire planes are used to detect the scintillation light produced by the particle interactions, providing a trigger signal used in coincidence with the accelerator beam signal to open the readout window. 

One of the physics goals of MicroBooNE is to detect the neutrino burst emitted by a nearby core-collapse supernova if it happens during the lifetime of the experiment. The burst spans an interval of tens of seconds, and with neutrino energies in the $5 - 50$~MeV range. The expected number of neutrino interactions in the MicroBooNE detector for a core-collapse supernova at 10~kpc is of the order of ten (based on the expectation for the DUNE Far Detector~\cite{Acciarri:2015uup}) detected through the reaction $\nu_{\rm e} + \rm{{}^{40}Ar} \to e^- + \rm{{}^{40}K^*}$. 
The detector is exposed to an intense cosmic-ray flux due to the near-surface location, which makes triggering on the few expected low-energy supernova neutrinos using the PMT system very challenging. Instead, MicroBooNE is pioneering a new approach for detecting supernova neutrinos in a LArTPC based on a continuous readout stream of the detector, which is implemented in parallel to the beam-driven trigger readout. An alert generated by the Supernova Early Warning System (SNEWS)~\cite{Antonioli:2004zb}, initially aimed for the astronomical community, is repurposed as delayed trigger. 
The data from the MicroBooNE detector is stored temporarily in hard-disk drives arranged as circular buffers for more than 48 h in normal conditions, awaiting an SNEWS alert to transfer the data to permanent storage. Even with the use of distributed computing, the data rates produced by the continuous readout of the TPC are too large and real-time compression is required, implemented as part of the detector readout.

\begin{figure}[htbp]
\centerline{\includegraphics[width=0.5\textwidth]{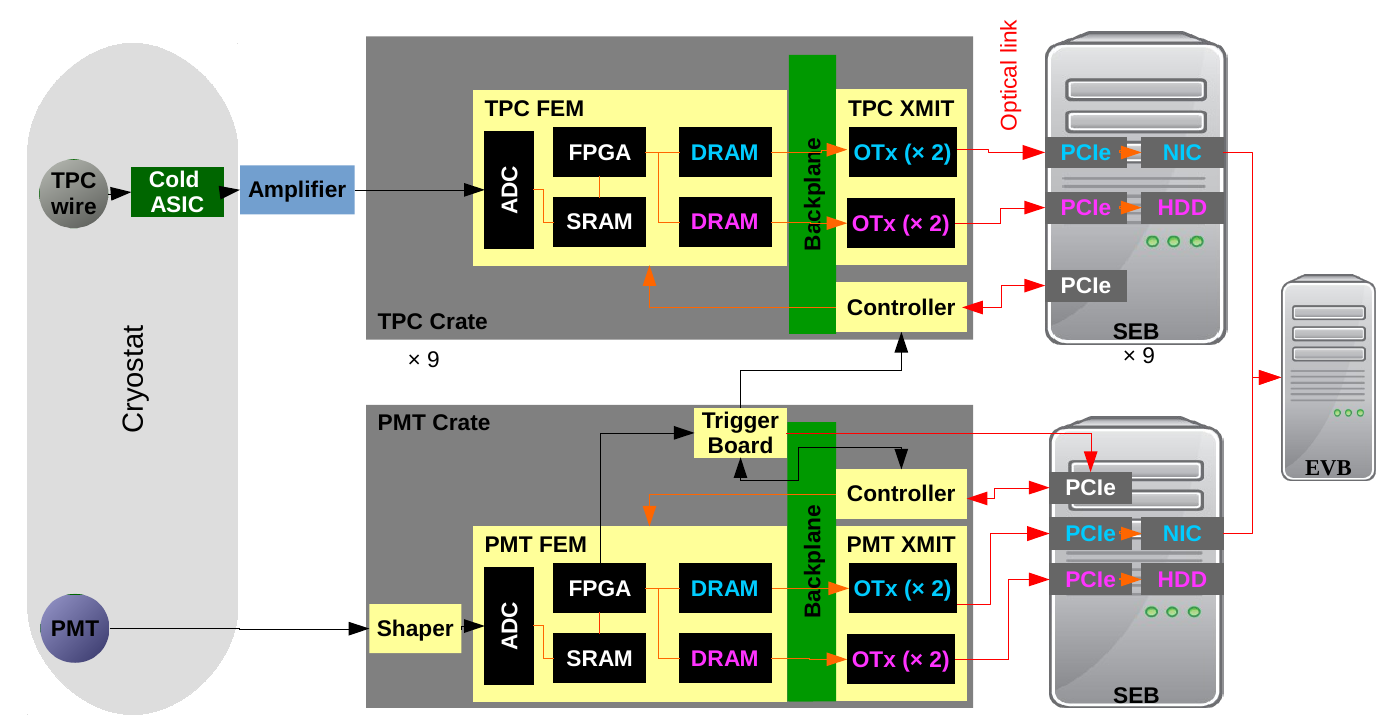}}
\caption{Diagram of the MicroBooNE detector readout. The Trigger Stream components are highlighted in blue text and the Continuous Readout Stream components are highlighted in magenta.}
\label{fig:ReadoutElectronics}
\end{figure}

The MicroBooNE readout electronics are shown in figure~\ref{fig:ReadoutElectronics}. The signals from the TPC wires are preamplified and shaped by ASICs submerged in the liquid argon, and then extracted from the cryostat via feedthroughs. The signals are amplified again by warm electronics sitting on top of the cryostat to condition them for transmission to the readout electronics using 20~m shielded twisted-pair cables.
The signals from 64 wires are read out by the Front-End Module (FEM), where they are digitized by eight 12-bit ADCs \cite{AD9222} at $16~\rm{MS/s}$. An Altera Stratix III FPGA \cite{StratixIII} downsamples the signals to $2~\rm{MS/s}$ and writes them in time order to a $1~\rm{M} \times 36~\rm{bit}$ $128~\rm{MHz}$ static RAM (SRAM). The SRAM is configured as a ring buffer, storing $12.8~\rm{ms}$ of data divided into 8 frames ($1.6~\rm{ms/frame}$). The FPGA reads out the TPC signals in channel order and splits them into two parallel streams: one is only read out upon trigger (Trigger Stream) and it is used for the physics program with the neutrino beam, another is continuously read out (Continuous Readout Stream) and it is used for non-beam physics such as supernova neutrinos (for this reason, it is also known as Supernova Stream). A transmitter board (XMIT) collects the data for each stream from each FEM through a VME crate backplane, with $512~\rm{MB/s}$ bandwidth, using a token-passing scheme. The backplane dataway is shared between both streams but the Trigger Stream is given priority. The FEM has a dynamic RAM (DRAM) for each stream to store the data awaiting its turn to be transferred.
The XMIT sends the data to custom PCIe $\times 4$ cards in the Sub-Event Buffer (SEB) DAQ server using  4 optical transceivers, two for the Trigger Stream and two for the Continuous Stream, each rated to $3.125~\rm{Gbps}$. The readout of the TPC is approximately equally distributed among 9 crates, each connected to one SEB. The Trigger Stream data from each SEB is sent to the Event Builder (EVB) DAQ server using a network interface card (NIC), while the Continuous Stream is written to a $15~\rm{TB}$ local hard-disk drive (HDD) in each SEB, awaiting a SNEWS alert to be further transferred to offline storage. If no SNEWS alert is issued and the disk occupancy reaches $80\%$, the oldest data is permanently deleted until the occupancy falls below $70\%$. 
The bottleneck of the Continuous Stream is the HDD writing speed, which is conservatively limited to $\sim 50~\rm{MB/s}$. In order to reduce the input data rate of $\approx 4~\rm{GB/s}$ at each crate, a $\sim 80$ compression factor is required. 

\section{FPGA-based real-time compression}\label{FPGACompression}
The FPGA in the FEM applies two data reduction algorithms sequentially to reduce the Continuous Readout Stream input data rate. The first one is a lossy compression using zero suppression, the second one is lossless compression using a Huffman encoding.

\subsection{Zero suppression}
The goal of the zero suppression is to remove the samples of the TPC waveforms which do not contain particle-induced signals, which is feasible since these signals are sparse in MicroBooNE .
The zero suppression is executed by applying an amplitude threshold with respect to an estimated baseline, and the ADC samples not passing the threshold are eliminated.
The sign of the threshold condition can be positive (samples passing are greater than the sum of the baseline and threshold), negative (samples passing are smaller than the baseline minus the threshold) or either.
In order to better capture the waveform, some samples before the first one that passes the threshold (presamples) and after the last one that passes the threshold (postsamples) are also acquired. The collection of these samples is defined as a Region-Of-Interest (ROI). The numbers of presamples and postsamples are configurable for each FEM. All MicroBooNE FEMs use 7 presamples and 8 postsamples, the maximum values allowed by the FPGA firmware.

The amplitude threshold and sign are configurable for each channel. 
Two methods have been used to determine them.
For the first one, a single threshold was used for all the channels belonging to the same TPC plane, with a physics-driven motivation of separating the signal (dominated by cosmic-ray muons crossing the detector) from the noise. For this, using previous raw data from the Trigger Stream, a simulation of zero suppression was applied with a loose threshold, and ADC spectra of the maximum and minimum values found in each ROI were produced. For each plane, two peaks were observed, interpreted as the noise and signal distributions. The ADC threshold was placed in the valley between them.
For the second method, individual thresholds were found for each channel. Again, previous raw data from the Trigger Stream was used, but the focus was shifted from physics to a bandwidth-driven approach, with the thresholds chosen to suppress the central $98.5\%$ of interval of the ADC distribution. 

The baseline upon which zero suppression is executed can be dynamically estimated or loaded as a static value.
For the static configuration, the ADC value for each channel baseline is set to the mode of the ADC distribution from previous data taken with the (non-zero-suppressed) Trigger Stream.
For the dynamic configuration, the channel baseline is estimated in real time using 3 contiguous blocks of 64 samples each.
A rounded mean ADC value for each block, $\mu_i$, is computed by summing the ADC value of all 64 samples and then dropping the 6 least significant bits (equivalent to an integer division by $64$). 
A truncated ADC variance for each block, $\sigma^{2}_{i}$, is computed by summing the squared differences between the ADC value of each sample and the rounded mean computed, $\mu_i$, and then dropping the 6 least significant bits. 
If the absolute value of the difference between an ADC sample and the rounded mean is greater than or equal to 63, the value is fixed to 4095 to avoid arithmetic overflows.
To prevent using a baseline from a block containing an actual signal or a non-representative baseline fluctuation, the rounded mean and the truncated variance of each block are compared to the ones from the two neighboring blocks.
If the three rounded mean differences ($|\mu_{i+1} - \mu_{i}|$, $|\mu_{i} - \mu_{i-1}|$, $|\mu_{i+1} - \mu_{i-1}|$) and the three truncated variance differences ($|\sigma^{2}_{i+1} - \sigma^{2}_{i}|$, $|\sigma^{2}_{i} - \sigma^{2}_{i-1}|$, $|\sigma^{2}_{i+1} - \sigma^{2}_{i-1}|$) between blocks are within configurable tolerance values, the rounded mean of the central block, $\mu_i$, is taken as the new baseline and applied for zero suppression for ADC samples beginning after the third block. 
The dynamic baseline estimation is applied as a sliding window of three 64-sample blocks from the beginning of the run, dropping the oldest 64-sample block and adding a newer block. 
The baseline tolerance parameters (rounded mean and truncated variance differences) are configured per FEM. The FPGA does not generate data for a channel until the rounded mean and the truncated variance differences between blocks in that channel satisfy the tolerance conditions.

\begin{figure}[htbp]
\centerline{\includegraphics[width=0.5\textwidth]{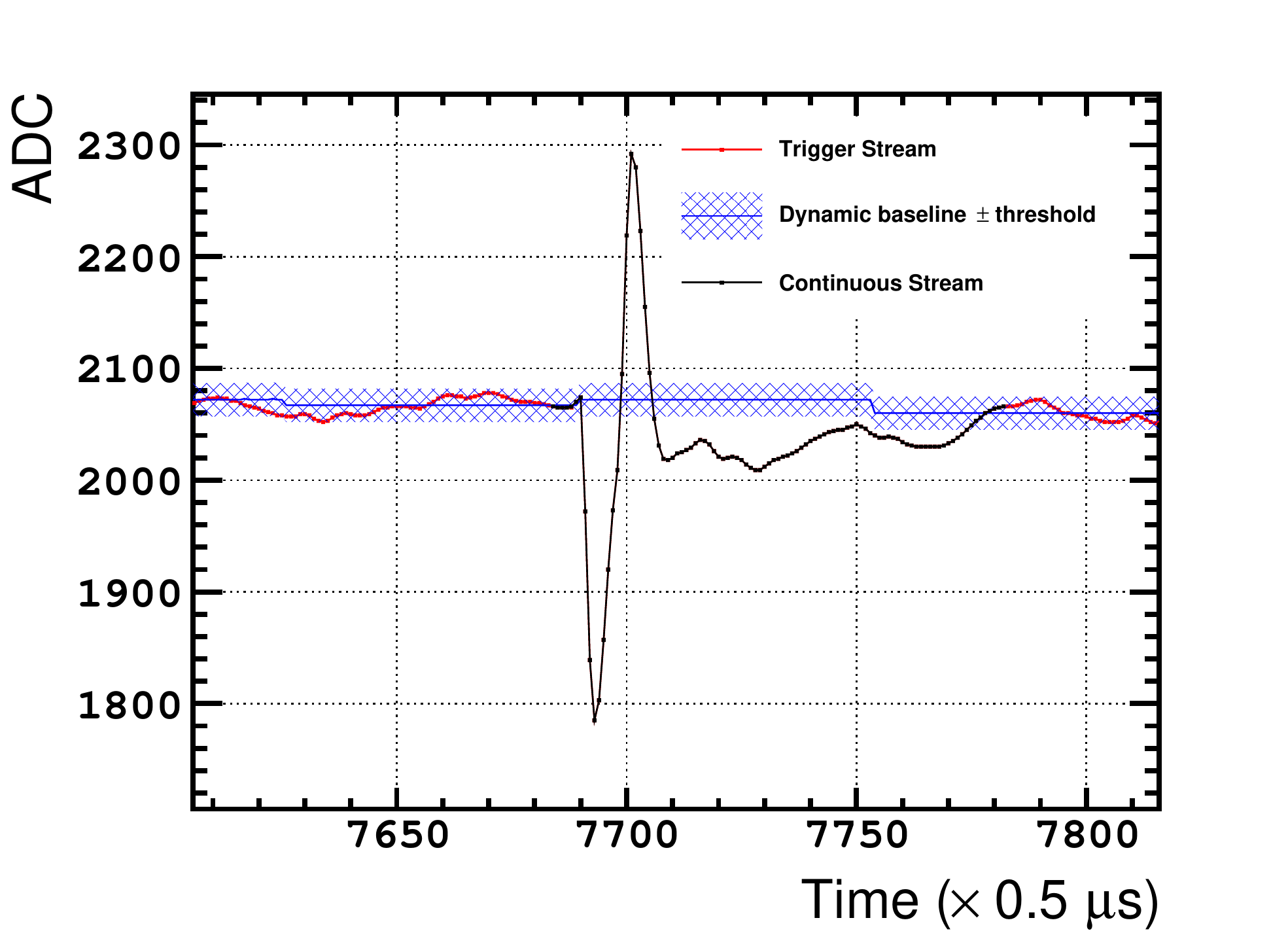}}
\caption{Example of data from the test stand at Nevis Laboratories showing a zero-suppressed waveform in the Continuous Readout Stream (black), superimposed on the same waveform from the Trigger Stream (red). An emulation of the dynamic baseline used by the FPGA is shown as a blue line, with the threshold shown as a cross-hatched band. Only the samples out of this band are saved, plus a number of samples preceding them (presamples) and following them (postsamples).}
\label{fig:ZeroSuppression}
\end{figure}

An illustration of the zero suppression with dynamic baseline is shown in figure \ref{fig:ZeroSuppression}.

\subsection{Huffman Encoding}
After the digitized waveform has been zero suppressed, it is run through a Huffman encoding.
The difference between successive samples is computed, $\mathrm{\Delta ADC = ADC_i - ADC_{i-1}}$, and if the absolute difference is smaller than 4 ADC counts, the sample value $\mathrm{ADC_i}$ is encoded using a fixed table (see table \ref{tab:Huffman}) relative to the preceding sample $\mathrm{ADC_{i-1}}$.

\begin{table}
\caption{Huffman encoding table}
\begin{center}
\begin{tabular}{|r|r|}
\hline
\multicolumn{1}{|c|}{$\mathbf{\Delta ADC}$} & \multicolumn{1}{|c|}{\textbf{Code}}\\
\hline
0 & 1\\
-1 & 01\\
+1 & 001\\
-2 & 0001\\
+2 & 00001\\
-3 & 000001\\
+3 & 0000001\\
\hline
\end{tabular}
\label{tab:Huffman}
\end{center}
\end{table}

The readout electronics data format uses 16-bit words. Non-Huffman-coded ADC samples are stored in the lowest 12 bits, and use the 4 upper bits as header to label the word. Huffman-coded words have the most significant bit set to 1 to label the word as Huffman-coded. The lower 15 bits are used to contain ADC information using the codes shown in table \ref{tab:Huffman}. 
If there are no more samples to be encoded in the Huffman word (e.g.\ the next ADC difference is larger than $\pm3$~ADC counts) or the code does not fit in the remaining available bits, the unused least significant bits are filled with zeros. In the latter case, a new Huffman-coded word is created to continue storing the ADC differences.

As an example, to code a portion of a waveform with relative ADC differences (-2, -1, 0, +1); the Huffman word is (the differences are written from right to left):
\begin{equation*}
\begin{split}
(\textrm{Huffman bit})(+1)(0)(-1)(-2)(\textrm{padding}) =\\
= 0\textrm{b}(1)(001)(1)(01)(0001)(00000)
\end{split}
\end{equation*}
In this example, the information of the four ADC values would have taken four 16-bit non-Huffman-coded words, but instead it is stored in a single 16-bit Huffman-coded word.

\subsection{Compression results}

\begin{figure}
\centerline{\includegraphics[width=0.5\textwidth]{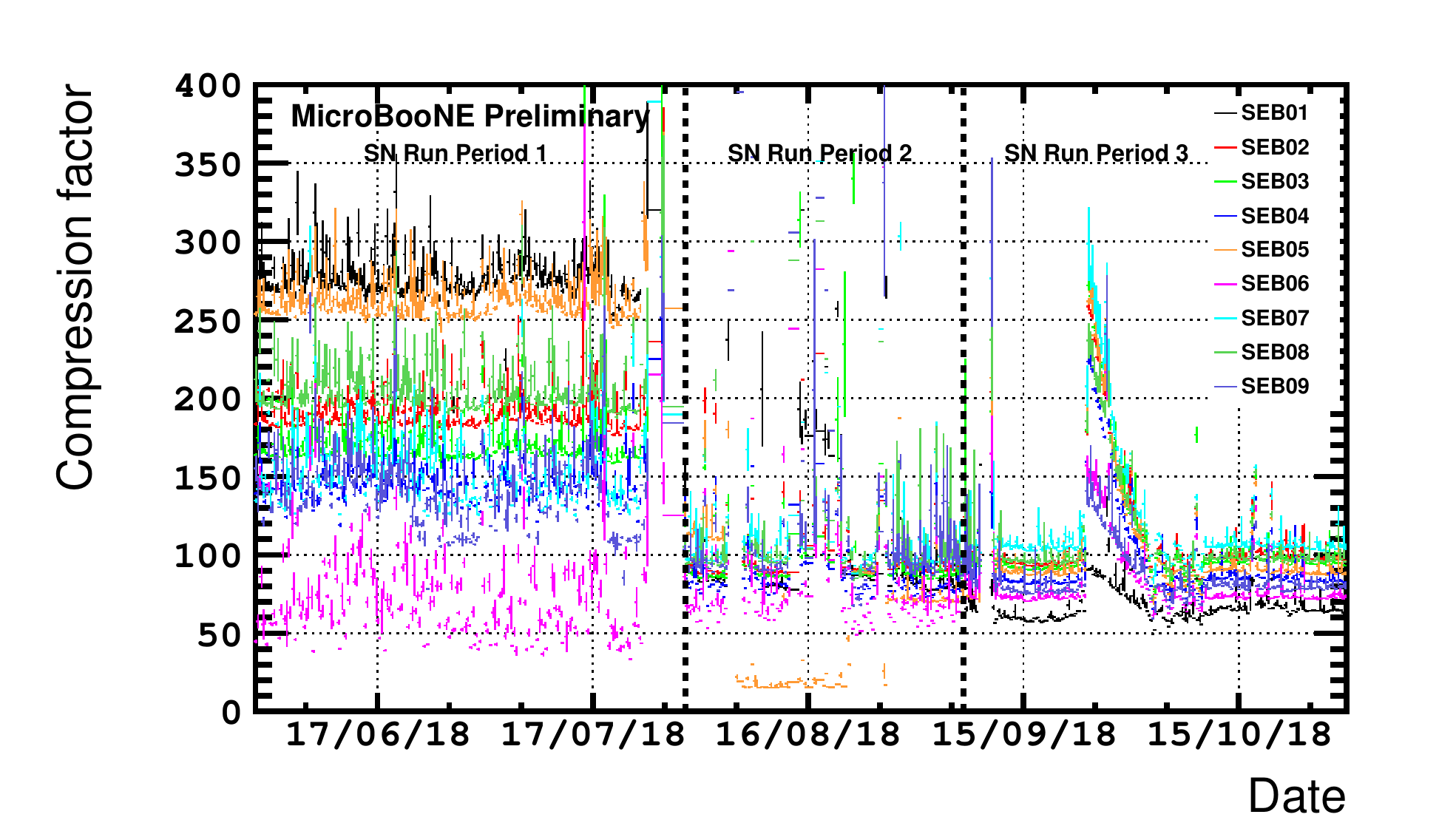}}
\caption{Compression factor achieved in the MicroBooNE Continuous Readout Stream for the 9 TPC DAQ servers (SEBs) with the three zero suppression configurations used so far. Date format is day/month/year. During SN Run Period 1 the configuration used plane-wise thresholds and dynamic baselines. During SN Run Period 2 the (mostly lower) channel-wise thresholds were tested, keeping the dynamic baseline estimation. The static baselines were introduced for SN Run Period 3 (September 2018 -- present). Each point shows the average data disk-write rate over 6 h. The low compression factor for SEB05 during mid-August was caused by a misconfigured ASIC after emerging from one power outage, and returned to the proper configuration in the following power outage. The ramp starting on September 24th, 2018 corresponds to the filling of the cryostat with a batch of lower-purity argon, followed by a period of HV instabilities.}
\label{fig:CompressionFactorQuery}
\end{figure}

Figure \ref{fig:CompressionFactorQuery} shows the compression factors measured in each SEB for the three zero-suppression configurations tested in the Continuous Readout Stream (table \ref{tab:ZSConfigs}). The compression factor is defined as the ratio of the expected data rate without compression and the measured data rate.

\begin{table}[htbp]
\caption{Zero-suppression configurations tested in MicroBooNE}
\begin{center}
\begin{tabular}{|c|c|c|}
\hline
\diagbox{\textbf{Thresholds}}{\textbf{Baseline}} & \textbf{Dynamic} & \textbf{Static}\\
\hline
\textbf{Physics-driven plane-wide} & SN Run Period 1& Not used\\
\textbf{Bandwidth-driven channel-wise} & SN Run Period 2& SN Run Period 3\\
\hline
\end{tabular}
\label{tab:ZSConfigs}
\end{center}
\end{table}

The physics-driven plane-wide thresholds with dynamic baseline were tested during SN Run Period 1 (November 2017 -- July 2018).
Compression factors well above the $\sim 80$ target were observed in all the SEBs except for SEB06, which showed higher data rates with large variations.
SEB06 reads out the majority of the noisy channels in the detector (due to misconfigured preamplifier/shaper ASICs), resulting in inaccurate estimations of the baseline position by the dynamic baseline algorithm. Other SEBs (e.g. SEB07 and SEB09) featured variations in data rates, but to a lesser extent.

The excessive compression observed in the SEBs during SN Run Period 1 suggested that the zero-suppression amplitude thresholds could be lowered to increase the efficiency for acquiring low-energy signals. This is especially important since zero-suppressed data cannot be recovered while a filter can be applied offline to reject noise that is acquired when lowering the thresholds. The bandwidth-driven channel-wise thresholds were tested during SN Run Period 2 (August 2018), keeping the dynamic baseline estimation. This resulted in lower thresholds for most of the channels, while the noisiest channels were effectively masked with high thresholds. This mitigated the excursions in data rates observed in SEB06, but the instabilities remained due to the baseline algorithm failing to find an accurate and stable baseline.

The dynamic baseline algorithm was replaced with a static baseline for SN Run Period 3 (September 2018 -- present). This allows to have a fixed baseline from the beginning of the run, required for the zero suppression, regardless of the channel noise levels. As a result, stable compression factors in the target region have been achieved.

\begin{figure}[htbp]
\centerline{\includegraphics[width=0.5\textwidth]{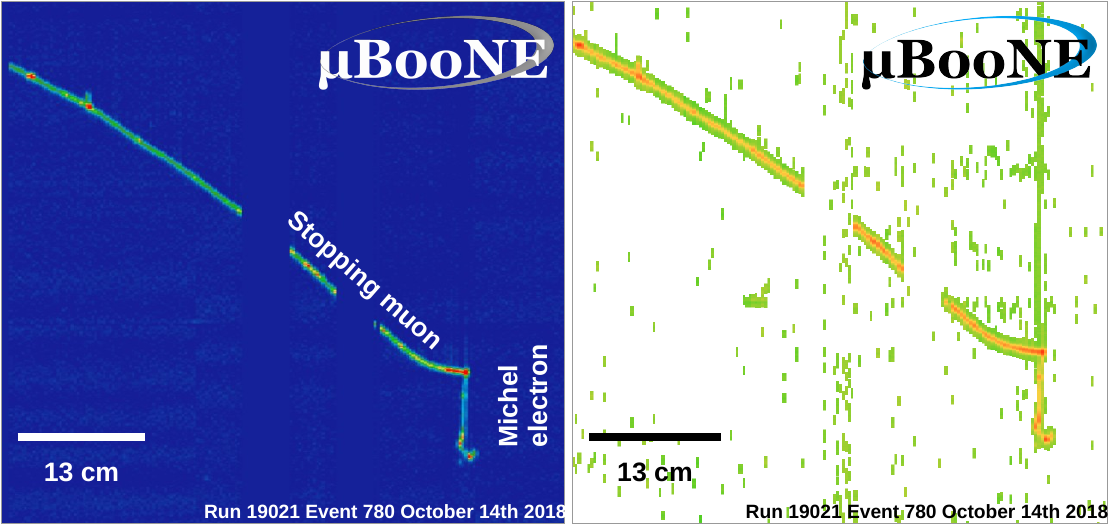}}
\caption{Stopping cosmic-ray muon decaying into a Michel electron. Left: original event from the Trigger Stream. Right: same event processed with the zero-suppression software emulation.}
\label{fig:MichelEventDisplays}
\end{figure}

\begin{figure}
\centerline{\includegraphics[clip, trim= 2cm 1.5cm 1cm 2cm,width=0.7\linewidth]{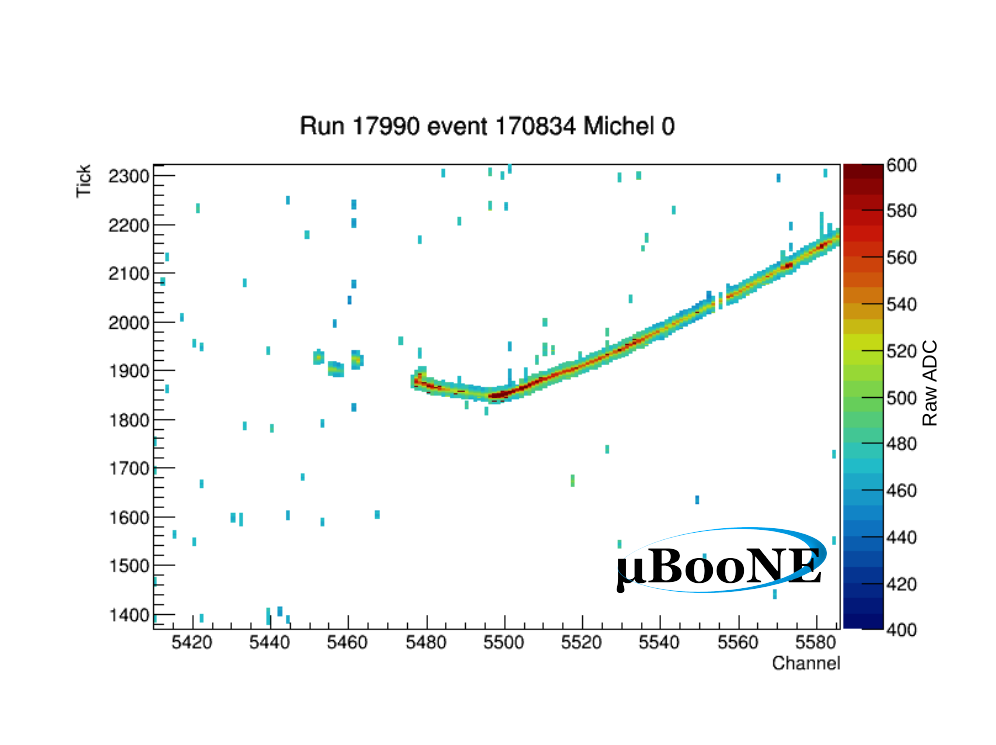}}
\caption{Example of a Michel electron candidate found in the Continuous Readout Stream on the collection plane. The horizontal axis shows the TPC channel number and the vertical axis shows the sample number (tick) in the 2 MHz clock. The color scale shows the uncalibrated ADC scale. The white background corresponds to the data which has been zero-suppressed.}
\label{fig:MichelCandidates}
\end{figure}

In order to demonstrate the sensitivity of the Continuous Readout Stream to supernova neutrinos, Michel electrons from stopping cosmic-ray muons are used, as they are similar to the ones expected from electron-neutrino interactions emitted by a core-collapse supernova. A comparison of the effect of (simulated) zero suppression on a Michel electron candidate is shown in figure \ref{fig:MichelEventDisplays}. Several thousands of Michel electron candidates have been found on the three TPC planes (see figure~\ref{fig:MichelCandidates}) using the same automated reconstruction and selection described in~\cite{Acciarri:2017sjy}, demonstrating the capability of the Continuous Readout Stream to acquire low-energy interactions.

\section{Conclusion}

MicroBooNE has achieved continuous readout of its liquid argon time projection chamber, enabling the acquisition of the neutrino burst of a nearby core-collapse supernova using the SNEWS alert as delayed trigger. The stream has been operated for almost two years, during which FPGA-based zero-suppression algorithms have been tested. The best results have been found using a combination of bandwidth-driven threshold and a static baseline for each channel, resulting in a stable $\sim 80$ compression factor. The sensitivity to supernova neutrino energies has been demonstrated by detecting and reconstructing low-energy electrons from stopping cosmic-ray muons on the three TPC planes.

\bibliographystyle{IEEEtran}
\bibliography{references} 

\end{document}